# Using Machine Learning and Big Data Analytics to Prioritize Outpatients in HetNets

Mohammed Hadi , Ahmed Lawey, Taisir El-Gorashi, Jaafar Elmirghani
*School of Electronic and Electrical Engineering*
*University of Leeds*
Leeds, United Kingdom
elmsha@leeds.ac.uk, a.q.lawey@leeds.ac.uk, T.E.H.Elgorashi@leeds.ac.uk, J.M.H.Elmirghani@leeds.ac.uk

*Abstract*— In this paper, we introduce machine learning approaches that are used to prioritize outpatients (OP) according to their current health state, resulting in self-optimizing heterogeneous networks (HetNet) that intelligently adapt according to users' needs. We use a naïve Bayesian classifier to analyze data acquired from OPs' medical records, alongside data from medical Internet of Things (IoT) sensors that provide the current state of the OP. We use this machine learning algorithm to calculate the likelihood of a life-threatening medical condition, in this case an imminent stroke. An OP is assigned high-powered resource blocks (RBs) according to the seriousness of their current health state, enabling them to remain connected and send their critical data to the designated medical facility with minimal delay. Using a mixed integer linear programming formulation (MILP), we present two approaches to optimizing the uplink side of a HetNet in terms of user-RB assignment: a Weighted Sum Rate Maximization (WSRMax) approach and a Proportional Fairness (PF) approach. Using these approaches, we illustrate the utility of the proposed system in terms of providing reliable connectivity to medical IoT sensors, enabling the OPs to maintain the quality and speed of their connection. Moreover, we demonstrate how system response can change according to alterations in the OPs' medical conditions.

## I. INTRODUCTION

According to [1], about 795 thousand people in the United States of America (USA) suffer a stroke each year, equivalent to an average of 1.5 stroke episodes each minute. Moreover, statistics from England, Wales and Northern Ireland for 2016-2017 indicate that one third of stroke patients went to hospital unaware of the date and time their symptoms began [2]. The severity of this issue appears even starker when we learn that the average time from the start of symptoms until admission to a hospital equates to 7.5 hours, with an additional 55 minutes for door-to-needle time (the duration between arrival at the emergency department and administering an anesthetic). This can be placed in perspective by noting that a stroke patient loses 1.9 million neurons per minute before the treatment starts [2]. Given its role in people's lives, health care is a vital subject. Moreover, it is one of the crucial areas where big data analytics (BDA), and machine learning (ML) technologies can make a difference, owing to the plethora of data generated by all network-enabled medical devices and the increasing convenience of electronic health record (EHR) collection [3]. Such technologies can be used optimally to analyze daily routine, allergies, diet, genetic information and a patient's EHR, and produce an accurate diagnosis much more quickly than a medical personnel with a certain degree of expertise [4]. According to [5], unknown risk factors can be identified using BDA for acute coronary syndrome (ACS),

Brugada syndrome, and spontaneous coronary artery dissection. The Hospital for Sick Children (SickKids) in Toronto is a good example of a real-life application of the use of BDA for disease prediction and response. SickKids applied advanced analytics to data comprising vital signs collected by bedside monitoring sensors in order to detect life-threatening infections in infants. The system was able to detect potential signs of an infection up to 24 hours in advance [6]. BDA was also used by the Medical Centre at Columbia University to identify complications suffered from bleeding stroke caused by ruptured brain aneurysms. Through the use of physiological data, the diagnosis reported complications 48 hours in advance, affording the health professionals sufficient time to address the situation [6].

The subject of patient monitoring is heavily reliant on the existence of functioning, network-connected, Internet of Things (IoT) and medical sensors attached or close to the patient. These sensors in turn require a stable and reliable connection to send their data. A cellular connection is favored over both Wi-Fi and wired connections as it does not confine the user to a small area (i.e., mainly indoors). However, such a connection can experience fading and path loss, where the signal to interference plus noise ratio (SINR) level is so low that the connection is unreliable or cannot be established. A channel that is slowly fading may imply that the signal level is unsuitable at the instant(s) when a critical OP's health information must be transported instantly to the medical facility. In this paper, we envision a dual role for the OP's data. Along with diagnosis, it guides the network operator to the OPs with the most urgent needs in order that resources can be directed toward them. We argue that ensuring high-quality connectivity between the OP-linked peripherals and their medical provider represents an important step toward highly personalized e-healthcare-centric services and applications. Topics such as resource allocation, patient monitoring, disease risk prediction, and prioritization are popular in the literature. Nonetheless, providing an optimization model transforming HetNets to function in an OP-conscious manner by combining those four topics is, to the best of our knowledge, unique.

The proposed approaches have the objective of maximizing the overall system SINR, with several power and resource block (RB) assignment constraints governing its operation. However, OPs are prioritized in the assignment scheme through allocating RBs with power proportional to their medical condition (i.e., the stroke likelihood). The assessment of the OPs' medical condition is determined by a naïve Bayesian classifier. In this work, we extend our previous research in [7], where we considered an LTE network with a single current state of user prioritization. The main contributions of this paper are: (i) using an interdisciplinary approach to develop two multi-tier HetNet optimization models incorporating the concepts of priority, e-healthcare,

ML/BDA, and resource allocation; and (ii) further investigating the system response over seven different current states resulting in different priority levels granted to the OPs. A current state refers to a vector of several values acquired by medical and IoT sensors (e.g., total cholesterol and blood pressure) that we run through the classifier to determine stroke probability, as shown in Table I (B). The remainder of this paper is structured as follows. Section II explores the related work. Section III presents the proposed system and the mixed integer linear programming (MILP) formulation of the RB assignment optimization problem. Section IV presents and discusses the results. The paper concludes with Section V.

## II. RELATED WORK

BDA has been described by [8] as a next-generation tool providing an optimal solution for the trade-off problems of resource utilization, sharing and allocation in wireless communication networks. We investigated this description comprehensively and extended it to include the potential role of BDA in the design of wired and wireless networks in [9]. The authors in [10] proposed minimizing the delay between the request and assignment of resources to users in radio access network (RAN). They proposed a big data processing environment to process log, alarm and configuration files to identify both user and network behaviors. Characteristics extracted from a stream of data were used by [10] to dynamically allocate clusters of cloud resources.

As new technologies emerge, the concept of prioritizing health care data and patients is gaining momentum. Priority-based cross layer routing and medium access channel protocols for health care applications were proposed by the authors in [11]. The proposed approach was effective in ensuring high reliability and energy saving in a wireless body area network. The authors in [12] proposed a system for storing and processing sensor data for health care applications, taking the health care big data security requirement into account. The authors proposed a prediction model utilizing MapReduce to predict heart diseases.

Cardiovascular disease (CVD) prediction using naïve Bayesian classifier and other ML/BDA techniques has been comprehensively discussed in previous literature. The authors in [13] used the naïve Bayesian classifier to detect cardiovascular disease and identify its risk level for adults. The classifier was validated by a number of cardiologists where more than 80% of respondents agreed with the classifier's accuracy. According to [14], the naïve Bayesian classifier was also compared to Decision Trees, yielding superior prediction accuracy to its counterparts. It should be noted that we have mathematically programmed the naïve Bayesian classifier using MILP to work seamlessly with our proposed models.

## III. NETWORK OPTIMIZATION FRAMEWORK

In this section, we introduce the system model, before explaining the role of the naïve Bayesian classifier and all of the stages of data preparation the data set must undergo before it can be employed in the proposed model. This section concludes with the problem formulation, where we present the main mathematical equations in terms of objective functions and constraints.

### A. System Model

We consider a HetNet comprised of a macro base station (MBS) and two neighboring Pico base stations (PBS) operating in an urban environment with a range of 40-100 meters. We assume that the network employs a spectrum partitioning strategy [15], and accordingly MBS users are not interfering with PBS users, hence, we consider here the intra-tier interference caused by users operating within the PBS range The users are randomly scattered and fall within two categories: healthy (normal) users, and OPs as illustrated in Fig.1. Due to placing the users at random distances from the PBS, different power levels are received at the PBS from their user equipment (UEs). If a low signal to interference plus noise ratio (SINR) channel is assigned to the OP, the health care provider may not be notified and the response may not arrive in time. The goal is to allocate high-gain RBs to OPs proportional to the severity of their medical status (i.e., stroke likelihood) as calculated in a cloud-located BDA engine according to the steps shown in Fig. 2 (thus prioritizing the OPs over normal users). OPs with high SINR values have greater spectral efficiency for their connection, because spectral efficiency is directly proportional to throughput, and the OPs will be able to send their data faster, hence minimizing the delay.

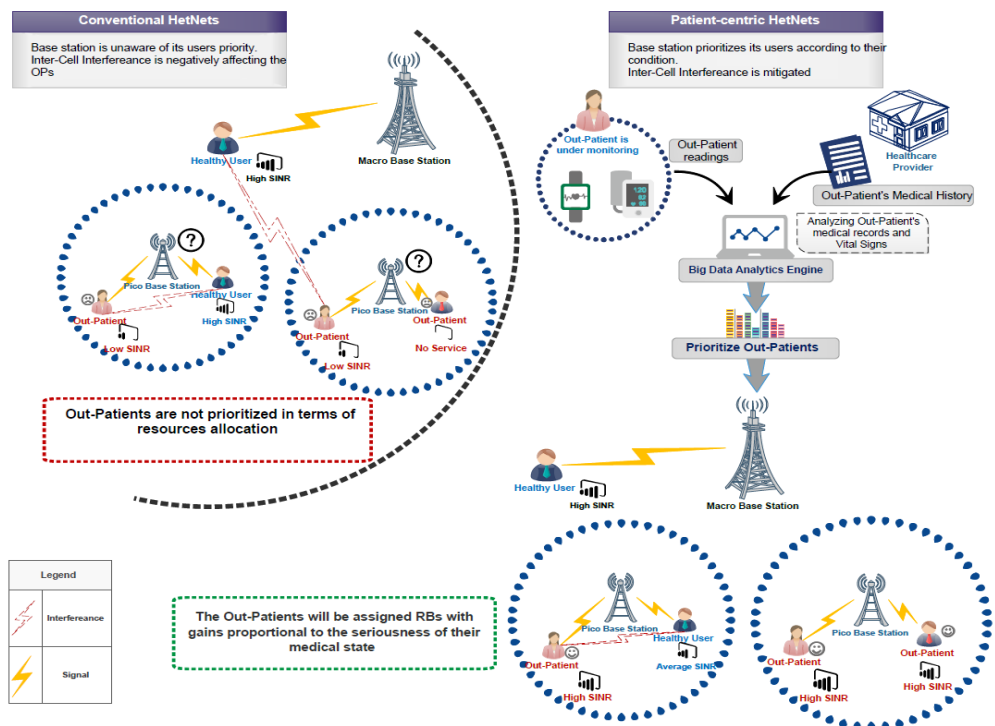

Fig.1. Patient-Aware HetNet

### B. Naïve Bayesian Classifier

The naïve Bayesian classifier is a probabilistic statistical classifier used in this work to determine the likelihood of a stroke. The classifier utilizes a number of independent feature variables $f_i$ (e.g., blood pressure and cholesterol levels) obtained from an historical data set (i.e., OP's medical record) to predict the likelihood of an incident $c$ (i.e., a stroke), as depicted in Fig. 2. The classifier is termed naïve because it assumes that feature variables are unrelated to one another [16]. We would like to highlight that further feature variables with the possibility of using the semi-naïve Bayesian classifier [17] or the locally weighted naïve Bayesian classifier [18] are being considered for future work. In addition to the classifier's track record in disease risk prediction (mentioned in Section II), we chose this classifier for the following reasons: (i) it exerts less computational burden due to its low complexity; (ii) the linearity of the classifier; (iii) it is optimal for any two-class concept with nominal features [19]; (iv) it does not require large training data sets [20]; and (v) it has proven accuracy in CVD prediction when compared with [21, 22]. Items (i) and (ii) are crucial in our research, given that the classifier is programmed jointly with the MILP.

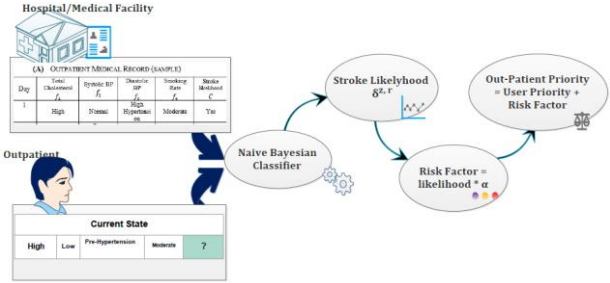

Fig.2. Outpatient Priority Calculation Procedure

The classifier's *posterior probability* is given as

$$\delta^{z,r} = p(C = c | F_i = f_i) = P(C = c) \prod_{i=1}^{n} P(Fi = fi | C = c) \quad (1)$$

where $P(C = c)$ represents *the prior probability* of stroke, $\delta^{z,r}$ denotes the stroke likelihood of OP $z$ ($Z \subset \mathcal{K}$) having class C take the $r^{th}$ value (in this case, r = yes), where $n$ represents the total number of feature variables, and the *likelihood* of $F$ given $C$ is given in (2)

$$p(F_i = f_i | C = c) = \frac{\sum_{i=1}^{n}(C = c \wedge F_i = f_i)}{\sum_{i=1}^{n}(C_i = C_i)} \quad (2)$$

where the term $\prod_{i=1}^{n} P(Fi = fi | C = c)$ represents the *joint probability*. Table I (A) represents a sample of the medical record of a single OP. Feature variables $f_1, \ldots f_4$ represent the main contributors to a stroke mentioned in [23, 24], and they are blood pressure (BP), cholesterol level, and smoking rate. We used the Framingham cardiovascular cohort study [25] to populate the data set of the individual OPs. It should be noted that the Framingham study contains readings for more than 3,000 persons. Due to privacy and regulatory reasons it was not possible for us to acquire cohort medical records for several *individual* patients. Thus, we segmented parts of the Framingham data set in [26] to represent several OPs.

### TABLE I

**(A) OUTPATIENT MEDICAL RECORD (SAMPLE)**

| Day | Total Cholesterol $f_1$ | Systolic BP $f_2$ | Diastolic BP $f_3$ | Smoking Rate $f_4$ | Stroke likelihood $C$ |
|---|---|---|---|---|---|
| 1 | High | Normal | High Hypertension | Moderate | Yes |
| 2 | Normal | Pre-hypertension | Low | Heavy | No |
| ⋮ | ⋮ | ⋮ | ⋮ | ⋮ | ⋮ |
| 30 | Optimal | High Hypertension | Pre-hypertension | Light | No |

**(B) (CURRENT STATE)**

| instance | $f_1$ | $f_2$ | $f_3$ | $f_4$ | $C$ |
|---|---|---|---|---|---|
| 1 | Normal | Pre-hypertension | Normal | Heavy | ? |

The ranges depicted in Table I (A) are based on those in Table II. To be as medically precise as possible, the discretized values of $f_1, \ldots f_3$ are in line with official organizations or governmental health institutes such as the American National Institute of Health and the British Stroke Association [27-29]. As for $f_4$, it is based on the ranges in [30].

### TABLE II

FEATURE VALUES AND THEIR CORRESPONDING LEVEL

| Feature | Range | Level |
|---|---|---|
| **Total cholesterol Level (mg/dl)** [27] | <200 | Optimal |
| | 200-239 | Normal |
| | 240+ | High |
| **Systolic BP (mmHg)** [28] [29] | <120 | Normal |
| | 120-139 | Pre-hypertension |
| | 140+ | High Hypertension |
| **Diastolic BP (mmHg)** [28] [29] | <80 | Normal |
| | 80-89 | Pre-hypertension |
| | 90+ | High Hypertension |
| **Smoking rate (Cig/Day)** [30] | 1 - 10 | Light |
| | 11 - 19 | Moderate |
| | 20+ | Heavy |

In order to bias the MILP so that OPs are assigned higher gain RBs, normal users are assigned a base user priority $UP_k$ of 1, while OPs are assigned the base weight *plus* another weight derived from the multiplication of the stroke likelihood $\delta^{z,r}$ by weight parameter $\alpha$ to give an effective yet reasonable priority.

$$\begin{aligned} UP_k &= 1 + \alpha \cdot \delta^{z,r} \\ \forall\ k &\in \mathcal{K}: z = k, k > NU \end{aligned} \quad (3)$$

where the OP's *updated* priority is given by (3). *NU* depicts the total number of normal users. In this work, we identified users 8-10 to be the OPs and users 1-7 to be the normal users.

Thus, $NU$ in this case equates to 7. Using different values of $\alpha$ effects the system response in terms of the users' SINR levels, as we shall illustrate in the results section.

*C. Problem Formulation and Model Parameters*

Using our experience in MILP optimization in [31-33], and physical layer modeling in [34-36], we developed a model to optimize RB allocation in HetNets using MILP. Our scenario is comprises a HetNet consisting of one MBS and two PBS. It is assumed that the network follows a spectrum partitioning strategy where Pico and macro users are on different RBs (i.e., mitigating uplink inter-cell interference). Hence, interference occurs among Pico users only. Consequently, $B$ PBSs are represented by the set $\mathcal{B} = \{1, ..., B\}$. Each PBS has a total of $N$ RBs depicted by the set $\mathcal{N} = \{1, ..., N\}$. A total of $K$ users, both normal and OPs, represented by the set $\mathcal{K} = \{1, ..., K\}$ are to be served in an instant of time by the PBSs using RB $n$ on PBS $b$. The target is to optimize the uplink of the network by maximizing the overall system SINR while prioritizing the OPs by allocating them high-gain RBs.

The SINR $\Psi_{k,n}^b$ of user $k$ connecting to PBS $b$ using RB $n$ is given as:

$$\Psi_{k,n}^b = \frac{\Omega_{k,n}^b X_{k,n}^b}{\Omega_{m,n}^b X_{m,n}^w + \sigma_{k,n}^b} \quad (4)$$

The numerator depicts the *signal* part of the equation, whereas the denominator consists of two parts, *interference* received from users connected to other PBSs on the same RB calculated as $\Omega_{m,n}^b X_{m,n}^w$ while the *noise* is represented by $\sigma_{k,n}^b$. $X_{k,n}^b$ is a binary variable equal to 1 when user $k$ is connected to the PBS $b$ using RB $n$; $m, m \neq k$ and $w, w \neq b$ denote the interfering user(s) and interfering PBS(s), respectively. However, in our case there is a single interfering PBS. Thus the objective function for the first model is given as:

**Objective:** Maximize

$$\sum_{k \in \mathcal{K}} \sum_{n \in \mathcal{N}} \sum_{b \in \mathcal{B}} \Psi_{k,n}^b UP_k \quad (5)$$

A focus on fairness among users is asserted in the second model. Therefore, the objective function becomes:

**Objective**: Maximize

$$\sum_{k \in \mathcal{K}} \sum_{n \in \mathcal{N}} \sum_{b \in \mathcal{B}} \ln \Psi_{k,n}^b \quad (6)$$

The logarithmic sum of *all* users' SINRs (i.e., before prioritizing the OPs) is maximized in the objective in (6). Fairness among users is thus achieved, albeit at the expense of the SINR values due to the natural logarithm's characteristics.

**Objective:** Maximize

$$\sum_{k \in K, 1 \leq k \leq NU} \ln \Psi_{k,n}^b + \sum_{k \in K, k > NU} \Psi_{k,n}^b UP_k \quad (7)$$

The objective after prioritizing the OPs will be as in (7).

Both models obey a number of assignment and power constraints: (i) limiting the usage of the RB to one user only; (ii) a power constraint ensuring that the user cannot utilize more than its maximum permitted power per connection; and (iii) another assignment constraint guarantees the connectivity of all users by setting the minimum number of utilized RBs per user to one. In addition, (iv) several linearization constraints govern the multiplication of continuous and binary variables. However, the use of the natural log in the second model calls for piece-wise linearization to be used. The model parameters are illustrated in Table III.

IV. RESULTS AND DISCUSSION

In this section, we used the parameters in Table III for a scenario of a network employing a spectrum partitioning strategy. The results illustrate two approaches to identifying the resource allocation problem: the weighted sum rate maximization (WSRMax) and the proportional fairness (PF). The first approach targets the maximization of the weighted sum rate of all users' SINRs, with its objective in (5). The second, however, enforces fairness among users through its objectives in (6) and (7) by maximizing the logarithmic sum of the users' SINRs. The MILP optimization was performed using AMPL/CPLEX software running version 12.5 on a PC with 16 GB RAM and a core i5 CPU.

TABLE III
Model Parameters

| Parameter | Description |
|---|---|
| System bandwidth | 3 MHz |
| Total number of RBs | 15 |
| Channel Model | Path Loss [37] and Rayleigh fading [38] |
| Number of MBS | 1 |
| Number of PBS | 2 |
| Number of RB per MBS | 10 |
| Number of RBs per PBS | 5 |
| Number of users | 10 |
| Number of normal users ($NU$) | 7 |
| Number of OPs | 3 |
| AWGN ($\sigma_{k,n}^b$) | -162 dBm/Hz [37] |
| Distance between user $k$ and BS $b$ | (40 - 100) m |
| Maximum transmission power per connection | 23 dBm [37] |
| UE transmission power per RB | 17 dBm |
| Base (i.e. normal user priority) weight | 1 |
| Outpatient priority $UP_k$ calculation method | Naïve Bayesian classifier |
| OP observation period | 30 Days |
| Weight Parameter ($\alpha$) | 50, 500, and 1000 |

Furthermore, we considered seven different *current states* in terms of input feature variables, as displayed in Table IV. We run each model over all seven different *current states* for 400 data files each representing randomized users' locations (i.e., random received power levels at the PBSs in each data file) to simulate 400 instances, and showing the average SINR. The seven current states produce different probabilities of strokes. These probabilities, along with different weight parameter α values, will be reflected as different SINR levels as shown in Fig. 4.

TABLE IV
Outpatient Current States

| Instance | Features | | | | Class |
|---|---|---|---|---|---|
| | $f_1$ | $f_2$ | $f_3$ | $f_4$ | $C$ |
| 1 | Normal | Pre-hypertension | Normal | Heavy | ? |
| 2 | High | High Hypertension | Normal | Light | ? |
| 3 | Normal | High Hypertension | High Hypertension | Moderate | ? |
| 4 | High | High Hypertension | High Hypertension | Heavy | ? |
| 5 | Normal | High Hypertension | Pre-hypertension | Light | ? |
| 6 | Normal | High Hypertension | High Hypertension | Light | ? |
| 7 | High | High Hypertension | High Hypertension | Light | ? |

It should be noted that to simplify the SINR calculation, we converted all logarithmic units (i.e., dBm) into linear scale

(i.e., m Watt), hence the resulting average SINR values in Figures 3 and 4 are unit less.

### A. The WSRMax Approach

*1) Before Prioritizing the OPs*

In this scenario, all users have equal priority (i.e., $UP_k = 1$). The average SINR is 830 (i.e., around 29 dB). However, observing the OPs (i.e., users 8, 9, and 10) in Fig. 3 (a), one can note that they have comparable SINRs to other (healthy) users, and sometimes actually lower, such as when comparing OPs 8 and 9 to user 7.

*2) After Prioritizing the OPs*

The OPs were granted high-gain RBs according to their priority level. A negligible drop (0.3) in the average SINR is observed when selecting the weight parameter $\alpha = 50$. However, all OPs were granted above-average SINRs as shown in Fig. 4. (a), (b), and (c). The OPs' SINRs increase with a focus on the OP with the highest priority in each state; moreover, we can notice that for $\alpha \geq 500$ all OPs are assigned SINRs above the average, with 9% and 16% maximum SINR decrease when $\alpha = 500$ and 1000, respectively.

### B. The PF Approach

*1) Before Prioritizing the OPs*

The average SINR in this scenario is equal to 320 (around 25 dB) as illustrated in Fig. 3. (b). Users 9 and 10 are assigned less than the average SINR. A difference in the SINR levels can be observed between the two approaches. This is due to the use of the natural logarithm as well as the location of users with close proximity to the PBS. When compared with the results in [7], we can clearly observe that the effect of the log differs. However converting the SINRs to their logarithmic form (i.e., dB) shows that the SINR is still within the optimal range of operation.

*2) After Prioritizing the OPs*

In this scenario, the system's average SINR has increased due to the fact that only the normal users remain subjected to the logarithmic function. On the other hand, the OPs have high SINR levels, as shown in Fig. 4. (d), (e), and (f). It should be noted that the effect of the increase of weight parameter $\alpha$ is minimal compared to the WSRMax approach.

## V. CONCLUDING REMARKS

This paper offers two multidisciplinary frameworks for patient-centric optimization of HetNets. A BDA/ML algorithm is embedded in resource allocation optimization and provides patient prioritization in the e-health setting studied. The target is to prioritize stroke outpatients in HetNets according to their current medical condition based on readings acquired from body-attached and nearby IoT sensors. As a result, the developed ML-driven resource allocation frameworks grant these patients high-gain RBs to ensure that they are always connected and can send their data with minimum delay. To that end, the WSRMax and PF approaches were presented and compared. The WSRMax approach maximizes the OPs' SINRs with less impact on normal users when compared to the PF approach. The PF approach maximizes the OPs' SINRs to a greater extent than the WSRMax approach, while a noticeable impact can be observed on normal users. With a false positive rate of 0.36, the current classifier can be further enhanced and compared to other algorithms to assess a patient's state, while the integration of more feature variables in a larger data set constitutes a basis for future work. Furthermore, investigating inter-cell interference as part of a larger model is currently being considered as a future direction.

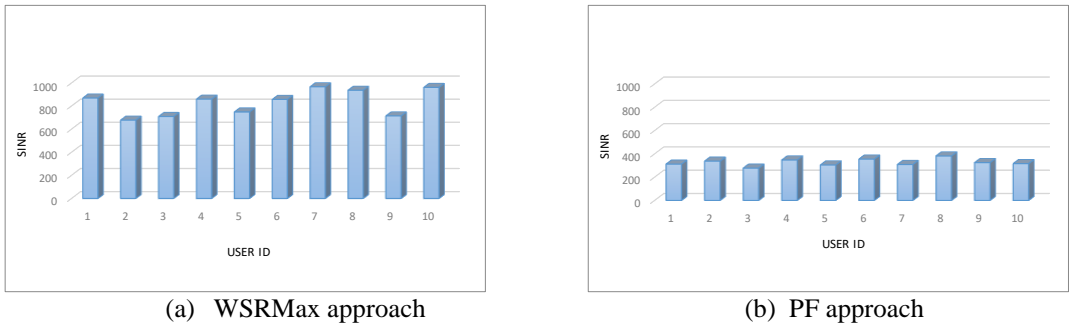

(a) WSRMax approach  (b) PF approach

Fig. 3. Users' SINRs before user prioritization

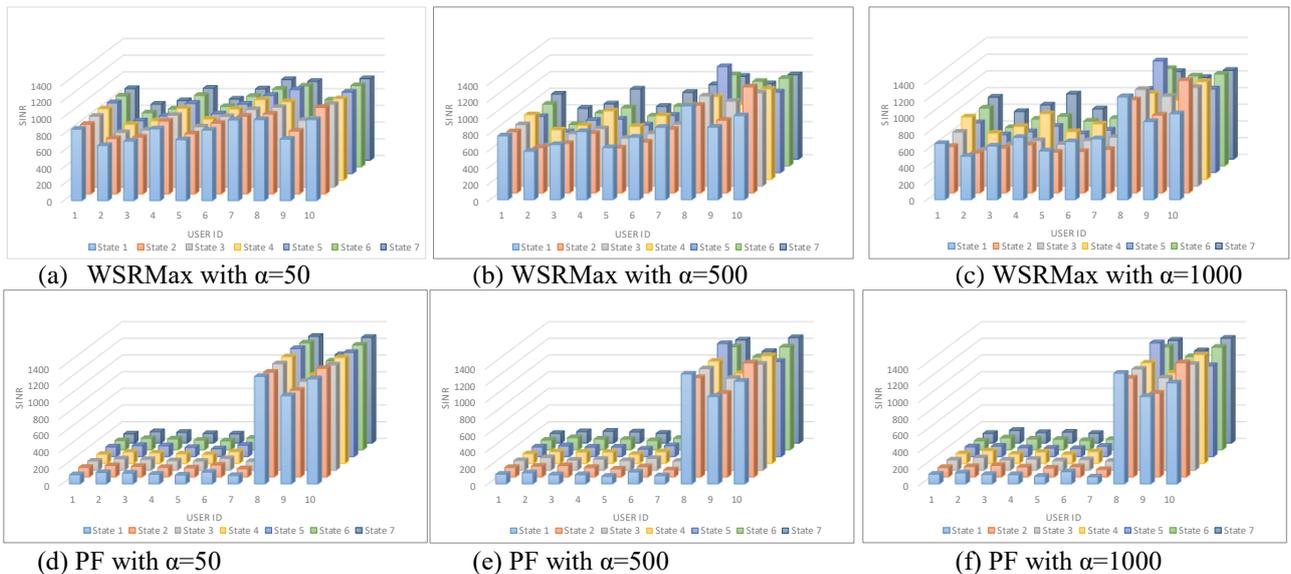

(a) WSRMax with α=50  (b) WSRMax with α=500  (c) WSRMax with α=1000

(d) PF with α=50  (e) PF with α=500  (f) PF with α=1000

Fig 4. Users' SINRs after user prioritization

## VI. ACKNOWLEDGMENT

The authors would like to acknowledge funding from the Engineering and Physical Sciences Research Council (EPSRC), INTERNET (EP/H040536/1) and STAR (EP/K016873/1) projects. All data are provided in full in the results section of this paper.

## VII. REFERENCES

[1] D. Mozaffarian et al., "AHA statistical Update," *Heart Dis. stroke*, vol. 132, 2015.
[2] S. Association, "State of the nation: Stroke statistics," *Retrieved June*, 2018.
[3] P. B. Jensen, L. J. Jensen, and S. J. N. R. G. Brunak, "Mining electronic health records: towards better research applications and clinical care," vol. 13, no. 6, p. 395, 2012.
[4] L. A. Winters-Miner, "Seven ways predictive analytics can improve healthcare," ed: Elsevier Connect, 2014.
[5] C. Krittanawong, H. Zhang, Z. Wang, M. Aydar, and T. J. J. o. t. A. C. o. C. Kitai, "Artificial intelligence in precision cardiovascular medicine," vol. 69, no. 21, pp. 2657-2664, 2017.
[6] W. Raghupathi and V. Raghupathi, "Big data analytics in healthcare: promise and potential," *Health information science and systems*, vol. 2, no. 1, p. 3, 2014.
[7] Mohammed Hadi, Ahmed Lawey, Taisir El-Gorashi, and a. J. Elmirghani, "Patient-Centric Cellular Networks Optimization using Big Data Analytics," *IEEE Access*, p. submitted for publication, 2018.
[8] P. Kiran, M. G. Jibukumar, and C. V. Premkumar, "Resource allocation optimization in LTE-A/5G networks using big data analytics," in *2016 International Conference on Information Networking (ICOIN)*, ed: IEEE, 2016, pp. 254-259.
[9] M. S. Hadi, A. Q. Lawey, T. E. El-Gorashi, and J. M. Elmirghani, "Big Data Analytics for Wireless and Wired Network Design: A Survey," *Computer Networks*, 2018.
[10] N. Kaur and S. K. Sood, "Dynamic resource allocation for big data streams based on data characteristics (5Vs)," vol. 27, no. 4, p. e1978, 2017.
[11] H. B. Elhadj, J. Elias, L. Chaari, and L. J. A. H. N. Kamoun, "A priority based cross layer routing protocol for healthcare applications," vol. 42, pp. 1-18, 2016.
[12] G. Manogaran, R. Varatharajan, D. Lopez, P. M. Kumar, R. Sundarasekar, and C. J. F. G. C. S. Thota, "A new architecture of Internet of Things and big data ecosystem for secured smart healthcare monitoring and alerting system," vol. 82, pp. 375-387, 2018.
[13] E. Miranda, E. Irwansyah, A. Y. Amelga, M. M. Maribondang, and M. Salim, "Detection of Cardiovascular Disease Risk's Level for Adults Using Naive Bayes Classifier," *Healthcare Informatics Research*, vol. 22, no. 3, pp. 196-205, 2016.
[14] M. Shouman, T. Turner, and R. Stocker, "Using data mining techniques in heart disease diagnosis and treatment," in *Electronics, Communications and Computers (JEC-ECC), 2012 Japan-Egypt Conference on*, 2012, pp. 173-177: IEEE.
[15] V. Chandrasekhar and J. G. J. I. T. o. C. Andrews, "Spectrum allocation in tiered cellular networks," vol. 57, no. 10, 2009.
[16] T. M. Mitchell, *Machine Learning*. McGraw-Hill Science/Engineering/Math, 1997.
[17] I. Kononenko, "Semi-naive Bayesian classifier," in *European Working Session on Learning*, 1991, pp. 206-219: Springer.
[18] E. Frank, M. Hall, and B. Pfahringer, "Locally weighted naive bayes," in *Proceedings of the Nineteenth conference on Uncertainty in Artificial Intelligence*, 2002, pp. 249-256: Morgan Kaufmann Publishers Inc.
[19] L. A. Muhammed, "Using data mining technique to diagnosis heart disease," in *Statistics in Science, Business, and Engineering (ICSSBE), 2012 International Conference on*, 2012, pp. 1-3: IEEE.
[20] M. Gandhi and S. N. Singh, "Predictions in heart disease using techniques of data mining," in *Futuristic Trends on Computational Analysis and Knowledge Management (ABLAZE), 2015 International Conference on*, 2015, pp. 520-525: IEEE.
[21] J. Butler and A. Kalogeropoulos, "Hospital strategies to reduce heart failure readmissions: where is the evidence?," ed: Journal of the American College of Cardiology, 2012.
[22] J. Soni, U. Ansari, D. Sharma, and S. Soni, "Predictive data mining for medical diagnosis: An overview of heart disease prediction," *International Journal of Computer Applications*, vol. 17, no. 8, pp. 43-48, 2011.
[23] A. Lewis and A. Segal, "Hyperlipidemia and primary prevention of stroke: does risk factor identification and reduction really work?," *Current atherosclerosis reports*, vol. 12, no. 4, pp. 225-229, 2010.
[24] M. L. Dyken, "Stroke risk factors," in *Prevention of stroke*: Springer, 1991, pp. 83-101.
[25] Framingham Heart Study. (26 April). *History of Framingham Heart Study*. Available: http://www.framinghamheartstudy.org/about-fhs/history.php
[26] W. W. LaMorte, "Using Spreadsheets in Public Health," *handout, School of Public Health, Boston University*, Handout accessed April 25, 2017.
[27] U. D. o. Health, H. Services, N. I. o. Health, L. National Heart, and B. Institute, "Your guide to lowering your cholesterol with TLC," *NIH Publication*, no. 06-5235, 2005.
[28] N. I. o. Health, *Your guide to lowering your blood pressure with DASH*. Smashbooks, 2006.
[29] S. Association, *Blood pressure information pack*. 2017.
[30] S. H. Jee, I. Suh, I. S. Kim, and L. J. Appel, "Smoking and atherosclerotic cardiovascular disease in men with low levels of serum cholesterol: the Korea Medical Insurance Corporation Study," *Jama*, vol. 282, no. 22, pp. 2149-2155, 1999.
[31] A. M. Al-Salim, T. El-Gorashi, A. Lawey, and J. Elmirghani, "Energy efficient big data networks: impact of volume and variety," *J. Trans. Netw. Serv. Manag.(TNSM)*, 2017.
[32] A. Q. Lawey, T. E. El-Gorashi, and J. M. Elmirghani, "Distributed energy efficient clouds over core networks," *Journal of Lightwave Technology*, vol. 32, no. 7, pp. 1261-1281, 2014.
[33] L. Nonde, T. E. El-Gorashi, and J. M. Elmirghani, "Energy efficient virtual network embedding for cloud networks," *Journal of Lightwave Technology*, vol. 33, no. 9, pp. 1828-1849, 2015.
[34] M. Hafeez and J. M. Elmirghani, "Analysis of dynamic spectrum leasing for coded bi-directional communication," *IEEE Journal on Selected Areas in Communications*, vol. 30, no. 8, pp. 1500-1512, 2012.
[35] I. J. G. Zuazola et al., "Band-pass filter-like antenna validation in an ultra-wideband in-car wireless channel," *IET Communications*, vol. 9, no. 4, pp. 532-540, 2015.
[36] R. Ramirez-Gutierrez, L. Zhang, and J. Elmirghani, "Antenna beam pattern modulation with lattice-reduction-aided detection," *IEEE Transactions on Vehicular Technology*, vol. 65, no. 4, pp. 2007-2015, 2016.
[37] J. P. Muñoz-Gea, R. Aparicio-Pardo, H. Wehbe, G. Simon, and L. Nuaymi, "Optimization framework for uplink video transmission in hetnets," in *Proceedings of Workshop on Mobile Video Delivery*, 2014, p. 6: ACM.
[38] P. Adasme, J. Leung, and A. Lisser, "Resource allocation in uplink wireless multi-cell OFDMA networks," *Computer Standards & Interfaces*, vol. 44, pp. 274-289, 2016.